%% file: paper.tex
\documentclass[sigplan,10pt]{acmart}

\acmConference[arXiv'21]{arXiv Preprint}{February, 2021}{Virtual}
\acmYear{2021}
\acmISBN{} 
\acmDOI{} 
\startPage{1}

\setcopyright{none}

\usepackage[normalem]{ulem}
\usepackage{pifont}
\usepackage{multirow}
\usepackage{fancyhdr}
\usepackage{setspace}
\usepackage{epsfig}
\usepackage{graphicx}
\usepackage{amsmath}
\usepackage{subfig}
\usepackage{color}
\usepackage{listings}
\usepackage{cprotect}
\usepackage{booktabs}
\usepackage{xcolor}
\usepackage{adjustbox}
\usepackage{color,soul}
\usepackage{float}
\usepackage{balance}
\usepackage{array}
\usepackage[lined, ruled]{algorithm2e}
\usepackage{enumitem}
\usepackage{siunitx}
\makeatletter
\newcommand{\removelatexerror}{\let\@latex@error\@gobble}
\makeatother
\newcommand{\cmark}{{\color{black}\ding{51}}}%

\graphicspath{{./Figs/}}

\begin{document}

\author{Angelo Matni }
\affiliation {
  \institution{Northwestern University}
}
\email{angelomatni2018@u.northwestern.edu}
\author{Enrico Armenio Deiana }
\affiliation {
  \institution{Northwestern University}
}
\email{EnricoDeiana2020@u.northwestern.edu}
\author{Yian Su }
\affiliation {
  \institution{Northwestern University}
}
\email{yiansu2018@u.northwestern.edu}
\author{Lukas Gross }
\affiliation {
  \institution{Northwestern University}
}
\email{lukasgross@u.northwestern.edu}
\author{Souradip Ghosh }
\affiliation {
  \institution{Northwestern University}
}
\email{souradipghosh2021@u.northwestern.edu}
\author{Sotiris Apostolakis }
\affiliation {
  \institution{Princeton University}
}
\email{sa8@cs.princeton.edu}
\author{Ziyang Xu }
\affiliation {
  \institution{Princeton University}
}
\email{ziyangx@cs.princeton.edu}
\author{Zujun Tan }
\affiliation {
  \institution{Princeton University}
}
\email{zujunt@cs.princeton.edu}
\author{Ishita Chaturvedi }
\affiliation {
  \institution{Princeton University}
}
\email{ishitac@cs.princeton.edu}
\author{David I. August }
\affiliation {
  \institution{Princeton University}
}
\email{august@cs.princeton.edu}
\author{Simone Campanoni }
\affiliation {
  \institution{Northwestern University}
}
\email{simonec@eecs.northwestern.edu}

\title[NOELLE Offers Empowering LLVM Extensions]{
  \underline{N}OELLE \underline{O}ffers \underline{E}mpowering
  \underline{LL}VM \underline{E}xtensions
}

\date{\today}

\begin{abstract}
\input{Sections/abstract}
\end{abstract}

\maketitle

\input{Sections/sections}

\bibliographystyle{ACM-Reference-Format}
\balance
\bibliography{Bibliography/bib,Bibliography/noelle-related-llvm-projects,Bibliography/compilers}

\end{document}

%% file: Sections/abstract.tex
Modern and emerging architectures demand increasingly complex compiler analyses
and transformations. As the emphasis on compiler infrastructure moves beyond
support for peephole optimizations and the extraction of instruction-level
parallelism, they should support custom tools designed to meet these demands
with higher-level analysis-powered abstractions of wider program scope. This
paper introduces NOELLE, a robust open-source domain-independent compilation
layer built upon LLVM providing this support. NOELLE is modular and
demand-driven, making it easy-to-extend and adaptable to custom-tool-specific
needs without unduly wasting compile time and memory. This paper shows the
power of NOELLE by presenting a diverse set of ten custom tools built upon it,
with a 33.2\% to 99.2\% reduction in code size (LoC) compared to their
counterparts without NOELLE.

%% file: Sections/sections.tex
\section{Introduction}
\label{sec:intro}
\input{Sections/introduction}

\section{NOELLE and Its Abstractions}
\label{sec:abstractions}
\input{Sections/noelle}

\section{Transformations Built Upon NOELLE}
\label{sec:clients}
\input{Sections/clients}

\section{Evaluation}
\label{sec:evaluation}
\input{Sections/evaluation}

\section{Related Work}
\label{sec:related_works}
\input{Sections/related_work}

\section{Conclusion}
\label{sec:conclusion}
\input{Sections/conclusion}

%% file: Sections/introduction.tex
The compiler community is on the front lines to satisfy the continuous demand
for computational performance and energy efficiency.
The focus of compiler advancements is shifting beyond peephole optimizations and
the extraction of instruction-level parallelism.
More aggressive optimizations and more sophisticated analyses with wider
scope are required to accommodate the needs of emerging architectures and
applications.

%

%
Modern compilers use low-level intermediate representations (IR) to
perform optimizations that are language-agnostic and
architecture-independent, such as LLVM IR from the LLVM compiler
framework~\cite{lattner2004llvm,::LLVMCompilerInfrastructure} and GIMPLE from
GCC~\cite{::GCCGNUCompiler}.
Low-level IR, along with a set of low-level abstractions built around it, is
designed to aid program analyses and optimizations and has shown its value for peephole optimizations and extraction of ILP. However, low-level abstractions
are not enough for advanced code analyses and transformations.
Consider automatic parallelization, one of the most powerful program
optimization techniques, exists only in a basic form~\cite{microsoft:msvc,
icc::AutomaticParallelization,::AutoParInGCCGCCWiki}, or does not exist at
all in most general-purpose compilers.
This paper shows that with proper abstractions, a daunting automatic
parallelization transformation can be implemented in fewer than a thousand
lines of code.

%

Advanced code analyses and
transformations go hand in hand with higher-level abstractions, as shown by
many existing compilers or frameworks. Several compiler infrastructures that
support automatic parallelization~\cite{::RoseCompilerProgram,
blume::PolarisNextGeneration, ::CetusProject} all operate on high-level
abstractions and perform source-to-source translation.
The recent success of domain-specific compilers/frameworks also proves the
importance of high-level abstractions for optimizations by uncovering
optimization opportunities at a domain-specific graph or operator
level~\cite{::TensorFlow,::Halide}.
However, these compilers limit themselves to specific program languages or
problem domains, and miss opportunities only presented in
low-level IRs, including more fine-grained operations and more canonical code
patterns.

The combination of higher-level abstractions and lower-level IR is the key to
advanced program analysis and optimizations. The claim can be found in the
SUIF compiler~\cite{::SUIFCompilerSUIF}, which provides low-level IR as well
as higher-level constructs~\cite{wilson::SUIFCompilerSystem}; and the IMPACT
compiler~\cite{chong::IMPACTArchitecturalFramework}, which provides
hierarchical IRs to enable optimizations at different levels. Despite the
claim, we are not aware of actively-maintained domain-independent compilers
that fulfill this combination.

While LLVM has become the de-facto compiler infrastructure to build upon, it
does not provide proper abstractions for advanced analyses and
transformations, including abstractions designed to describe properties of a
wider code scope (e.g., program dependence graph, program call graph) or abstractions that provide the mechanisms for advanced code
transformations (e.g., loop transformations, code scheduler, task creation).
These abstractions can ease the implementation of new transformations and
make the existing code transformations available in LLVM more powerful.





We propose a new open-source compilation layer called NOELLE that brings
forth abstractions for advanced code analyses and transformations. To
demonstrate the importance of NOELLE, we have implemented ten advanced code
transformations, nine of which need only a few lines of code. Only one of these transformations is
already available in LLVM (loop invariant code motion). We will show that
our version is significantly more powerful, requires significantly fewer
lines of code, and the implementation is more elegant than the LLVM
counterpart. The other nine transformations are missing in LLVM because they
are challenging to implement with the low-level abstractions LLVM
provides.

We have implemented a variety of code transformations upon NOELLE: a few
parallelizing compilers, a Pseudo-Random value generator selector, a
comparison optimization for timing speculative micro-architectures, a dead
function elimination, a memory guard optimization, a code analysis and
transformation to replace hardware interrupts, and a loop invariant code
motion. We call them NOELLE's custom tools. It is a challenge to implement
these custom tools only using the low-level abstractions provided by LLVM. Relying
on NOELLE, instead, most tools are implemented in only a few hundred lines
of code. We tested all these tools on 41 benchmarks from three benchmark
suites (SPEC CPU2017, PARSEC 3.0, and MiBench). All these tools improve the
quality of the code generated by LLVM with its highest level of optimization.
Finally, the high heterogeneity between these ten custom tools suggests NOELLE
provides general abstractions and support for a wide variety of advanced code
analyses and transformations.
Finally, we have released NOELLE publicly (https://github.com/scampanoni/noelle).

This paper:
\vspace{-0.13cm}
\begin{itemize}[leftmargin=*]

\item{\begin{sloppypar}introduces NOELLE, a robust open-source domain-independent compilation
layer built upon LLVM;\end{sloppypar}}

\item{describes the abstractions provided by NOELLE
(Section~\ref{sec:abstractions}) that ease the development of advanced code
transformations and analyses;}

\item{presents the tools provided by NOELLE (Section~\ref{sec:NOELLE_tools})
that ease the deployment of custom compilation tool-chains;}

\item{describes the testing infrastructure provided by NOELLE
(Section~\ref{sec:NOELLE_testing}) that enables automatic testing of
custom tools;}

\item{describes a diverse set of ten custom tools built upon NOELLE
(Section~\ref{sec:clients}) and highlights the
benefits of NOELLE's custom tools compared to vanilla LLVM
(Section~\ref{sec:evaluation-reduces});}

\item{evaluates the accuracy of NOELLE's abstractions (Section~\ref{sec:evaluation-abstractions}); and,}

\item{further motivates the need for NOELLE by comparing it with prior work
(Section~\ref{sec:related_works}).}

\end{itemize}

%% file: Sections/noelle.tex
Next, we describe NOELLE, its abstractions, and its tools.

\begin{figure}[htp]
\centering
\includegraphics[width=\columnwidth]{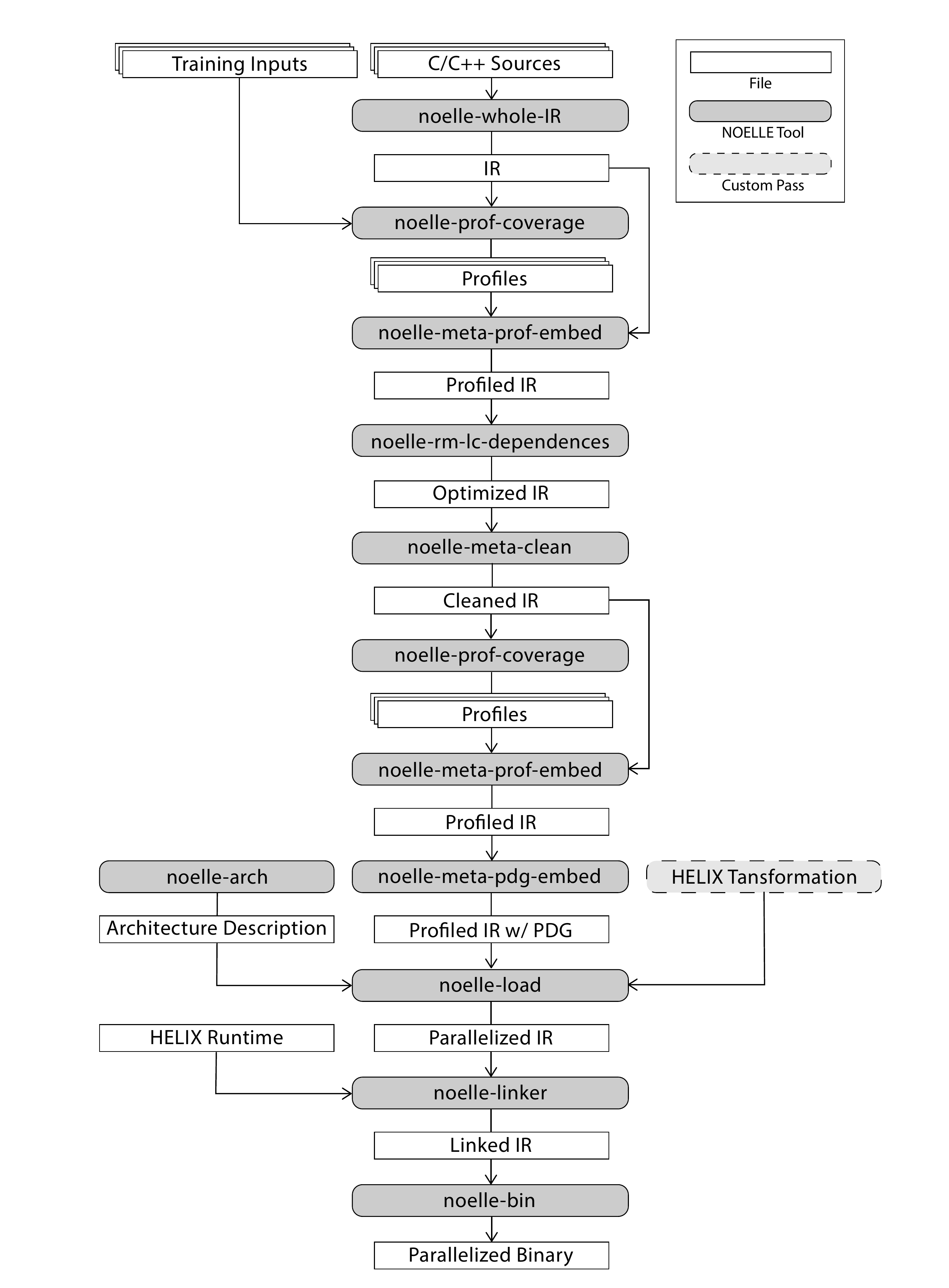}
\vspace*{-5mm}
\caption{
\label{fig:NOELLE_parallelization}
\small{
Compilation flow of the HELIX custom tool using NOELLE tools and a custom
pass, HELIX Transformation. Figure~\ref{fig:NOELLE}
shows in detail how to build HELIX Transformation using NOELLE abstractions.
}
\vspace{-0.3cm}
}
\end{figure}

\subsection{NOELLE in a Nutshell}
\label{sec:nutshell}
The goal of NOELLE is to provide abstractions that enable a simple implementation of complex code analyses and transformations (we call them custom tools) that target wide program scopes.
Custom tools built upon NOELLE include LLVM passes that work at the IR level to perform their code analyses and transformations.
Allowing these custom tools to be easily implementable and maintainable
requires simple and domain-independent abstractions powered by either
accurate low-level code analyses or complex low-level code transformations.
NOELLE provides such abstractions (Section~\ref{sec:abstractions}) with a modular design allowing its users to pay only the cost of creating the abstractions requested.

NOELLE's abstractions are powered by code analyses, some of which are provided by third parties.
For example, the PDG abstraction NOELLE provides is computed by running several alias analyses implemented by external codebases (SCAF~\cite{apostolakis:20:pldi} and SVF~\cite{sui:2016:SVFInterproceduralStatic}).
Moreover, NOELLE's modular design makes it easy to extend the list of external code analyses that power NOELLE's abstractions.
NOELLE also provides tools (Section~\ref{sec:NOELLE_tools}) for faster user-specific compilation flows.
Finally, NOELLE provides a testing infrastructure (Section~\ref{sec:NOELLE_testing}) to facilitate automatic testing of NOELLE itself as well as custom tools built upon it.

\paragraph{Input and Output}
The input of a compilation flow built upon NOELLE is the source code of a
program and optionally, a set of training inputs that could be used for profile-guided or autotuning-based custom tools.
The output is a binary for a target architecture supported by vanilla LLVM backends.

\paragraph{An Example of Compilation Flow}
NOELLE enables its users to deploy custom compilation flows by providing a set of tools, described in Section~\ref{sec:NOELLE_tools}.
Next, we describe an example of a compilation flow built using NOELLE's tools (shown in Figure~\ref{fig:NOELLE_parallelization}).
This is the compilation flow used by the custom tool HELIX (further described in Section~\ref{sec:clients}).

\begin{figure}[t]
\centering
\includegraphics[width=\columnwidth]{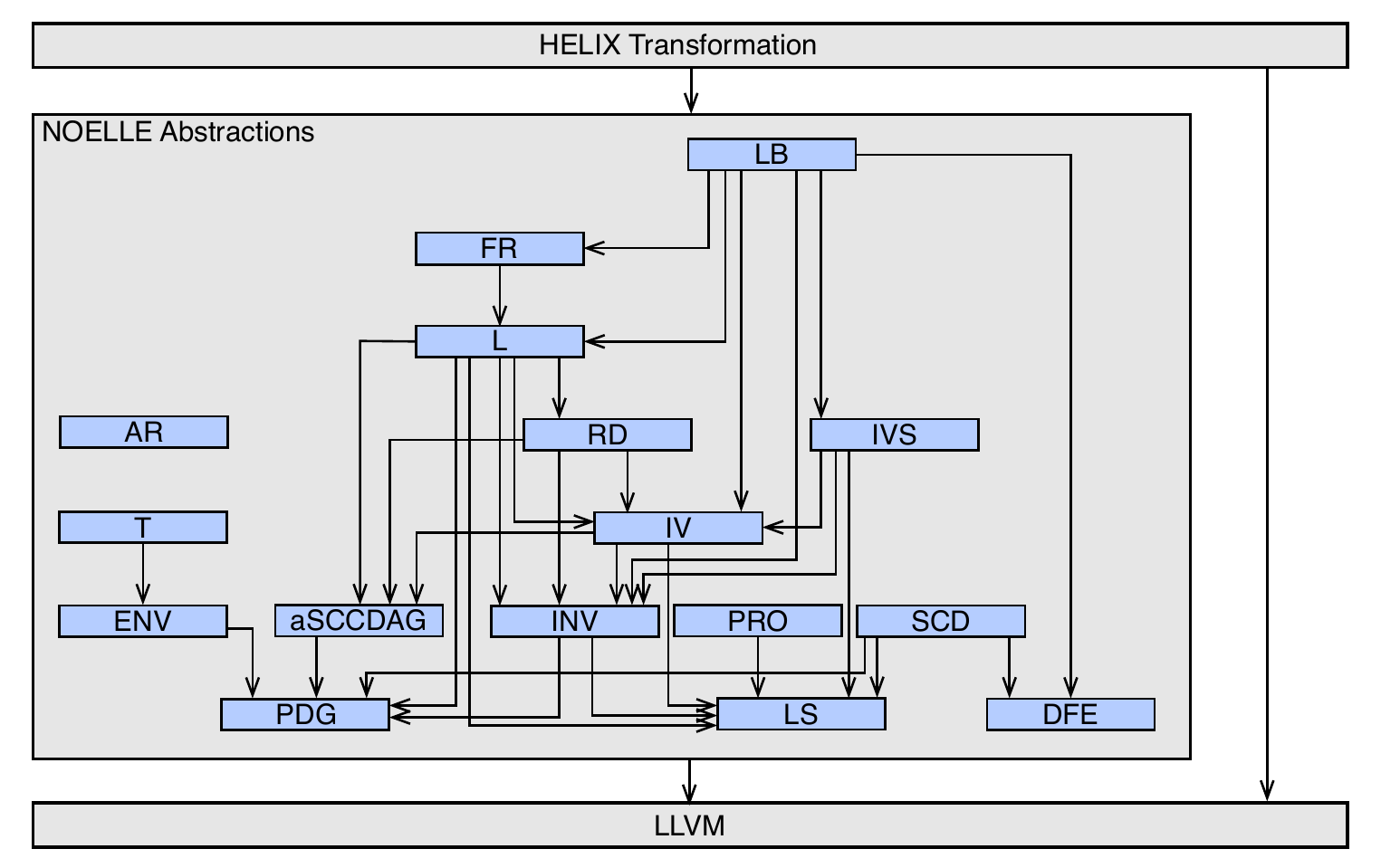}
\vspace*{-7mm}
\caption{
\label{fig:NOELLE}
\small{
HELIX transformation, a custom pass written using NOELLE abstractions. Arrows in the graph describe the dependence
between analyses. Refer to Table~\ref{tab:NOELLE_abs} for descriptions
of all NOELLE abstractions. Table~\ref{tab:client_abstractions} and
Table~\ref{tab:NOELLE_tools} describe abstractions used per custom and
NOELLE tool respectively.
}
\vspace{-0.3cm}
}
\end{figure}

Each source file composing a program being compiled is consumed by \texttt{noelle-whole-IR}, which outputs a single LLVM IR file that includes the whole program's code as well as options to use to generate the final binary (e.g., the libraries to link with).
Then, using traininig inputs given to NOELLE, \texttt{noelle-prof-coverage} runs several profilers to collect statistics about the single IR file's execution.
These statistics include the hotness of code regions (e.g., a loop, a basic
block), loop-specific information (e.g., the total number of iterations of
a loop, the average number of iterations per invocation of a loop), and
function-specific information (e.g., number of invocations of a function,
the average number of recursive calls of a recursive function).
The program's profiles are then embedded into the IR file by \texttt{noelle-meta-prof-embed}.
The generated IR is consumed by \texttt{noelle-rm-lc-dependences}, which applies a set of code transformations that aim to reduce loop-carried data dependences in hot loops (the minimum hotness required to consider a loop).
The generated IR is now more amenable to loop-centric code parallelization techniques.
The tool \texttt{noelle-meta-clean} cleans all NOELLE-specific metadata from the IR file.
Then, \texttt{noelle-prof-coverage} and the tool \texttt{noelle-meta-prof-embed} re-generate and embed the program's profiles, respectively.
Then, \texttt{noelle-meta-pdg-embed} computes the program dependence graph (PDG) and embeds it as metadata inside the IR file.
The \texttt{noelle-arch} computes architecture-specific profiles (e.g., communication latency between cores).
Its output is used by the HELIX transformation.
Finally, the \texttt{noelle-load} tool is invoked, which loads in memory NOELLE's compilation layer, to run the HELIX transformation.
The HELIX transformation relies on NOELLE's abstractions to parallelize hot loops.
The generated parallelized IR file is then consumed by \texttt{noelle-linker}, which embeds the HELIX-specific runtime into the IR.
Finally, \texttt{noelle-bin} generates the parallel binary.

\subsection{NOELLE's Abstractions}
\label{sec:abstractions}
Next, we describe the abstractions that NOELLE provides to its users.
NOELLE's abstractions (summarized by Table~\ref{tab:NOELLE_abs}) are demand-driven to preserve compilation time and memory.
Hence, users only pay for the abstractions they need.
In other words, if a user does not need the program dependence graph (PDG), then it will not pay the cost of analyzing the program to compute its dependences.

\input{Tables/NOELLE_abstractions}

\paragraph{PDG.}
NOELLE provides the Program Dependence Graph (PDG) representation of a program~\cite{ferrante1987program}.
This is obtained by extending NOELLE's \emph{dependence graph}, a templated class designed to represent a generic graph of directed dependences between nodes.
What constitutes a node is decided when the class is instantiated.
For example, the PDG instantiates this templated class with the LLVM instruction class.
Hence, the nodes of the PDG are the instructions of a program.
Each edge of the dependence graph contains attributes to differentiate between control and data dependences.
Data dependences are further characterized based on the dependence type
(Read-After-Write, Write-After-Write, Write-After-Read), whether it is
loop-carried, dependence distance, whether it is memory or register
dependence, and whether it is an apparent or actual dependence~\cite{Deiana:2018:UPN:3173162.3173181}.

An analysis or transformation (i.e., pass) built upon NOELLE can use the PDG abstraction to create loop dependence graphs and function dependence graphs.
The former is a dependence graph of a specific loop.
The latter refers to dependences only between the instructions of a function.
When a pass requests the loop dependence graph from a PDG, NOELLE runs loop-centric analyses to refine (and improve the precision about) the dependences that are included in the PDG for the specific loop in-question.

Users of this abstraction often want to know not only about the nodes of a
dependence graph that belong to a related code region (e.g., instructions
of a loop for a loop dependence graph) but also about the inputs, the
outputs, or both of the graph.
For example, a parallelizing code transformation of a loop needs to know the live-in and live-out sets of the target loop.
Because of this need, the templated class \emph{dependence graph} offers
two sets of nodes, the internal and the external ones.
The former belong to the related code region; the latter represents the
live-ins, live-outs, or both of that code region.
The computation of both sets of nodes is computed by NOELLE when a pass requests either a loop dependence graph or a function dependence graph.

\paragraph{aSCCDAG.}
Advanced code transformations like parallelization techniques can be implemented as different strategies to schedule instances of the nodes that compose the SCCDAG of a loop~\cite{tarjan1972depth, 1540952}.
For instance, HELIX distributes instances of a given SCCDAG node around the cores.
DSWP instead distributes nodes of an SCCDAG between cores while keeping all instances of a given node within the same core.
Hence, an important abstraction is the SCCDAG.
To this end, we introduce the \emph{augmented SCCDAG} abstraction or \emph{aSCCDAG}.
An aSCCDAG of a given loop is a complete description of loop dependences,
including those with the rest of the program.

A node of an aSCCDAG can be \texttt{Independent}, \texttt{Sequential}, or \texttt{Reducible}.
This categorization of a node $n$ depends on the relation between the
instructions' dynamic instances included in $n$ for a given loop invocation.
If all these instances are independent of each other, then $n$ is tagged as \texttt{Independent}.
If an instance of an instruction of $n$ depends on another instance of an instruction of $n$, then this node is tagged as \texttt{Sequential}.
Finally, if there are dependences between instances of $n$, but they are
reducible by a reduction code transformation (e.g., by cloning the defined
variable \texttt{s} in \texttt{s += work(d)}), then $n$ is tagged as
\texttt{Reducible}, and the related reduction is described within the node.

\paragraph{Call graph (CG).}
NOELLE provides the \emph{call graph} of a program where nodes are
functions, and edges indicate a given function invokes another.
This abstraction relies on the PDG to compute the possible callees of an indirect call.
Edges of the NOELLE's call graph can be must or may depend on whether a given caller-callee relation is proved to hold or not.
Each edge has sub-edges to indicates with which specific instructions a caller invokes another function.
Finally, CG can compute the set of disconnected islands of such a graph.

\input{Tables/NOELLE_tools}

NOELLE's call graph differentiates with LLVM's one by being complete: the
latter does not compute an indirect call's possible callees.
By being complete, NOELLE's call graph enables custom tools to assume that
the call graph's lack of an edge means a function cannot invoke another.
CG is used by the DeadFunctionEliminator custom tool built upon NOELLE,
aiming to reduce the binary size of a program.

\paragraph{Environment (ENV).}
NOELLE offers the \emph{Environment} abstraction, which is an array of pointers of variables.
Variables within an Environment represent the incoming and outgoing values from and to a set of instructions.
This set of instructions is described by a subset of the nodes of an aSCCDAG.
An example pass that relies on the Environment abstraction is a
parallelization technique that needs to propagate values explicitly between the cores.
Finally, NOELLE provides \emph{Environment Builder} to create, modify, and query environments.

\paragraph{Task (T).}
NOELLE offers the \emph{Task} abstraction to describe a code region that runs sequentially.
Parallelization techniques use the above abstraction in the following way.
Nodes within an aSCCDAG are partitioned into tasks.
An Environment is created for each task.
At runtime, tasks are submitted to a thread-pool, which will run them in parallel across the cores.
The explicit forwarding of data values between tasks is performed by loading/storing values from/to variables pointed by their environments.

\paragraph{Data flow engine (DFE).}
NOELLE provides a \emph{data flow engine} that can be used to implement data flow analyses.
DFE implements conventional optimizations like bitvectors, basic block granularity optimization, working list algorithm, and loop-based priority~\cite{appel2008modern}.
Finally, NOELLE provides a set of common data flow analyses that rely on DFE.

\paragraph{Profiler (PRO).}
NOELLE provides several code profilers, the ability to embed their results
into IR files, and abstractions to facilitate high-level queries on such data.
Examples of queries that can be performed are the hotness of a code region
(e.g., a loop, an SCC of a dependence graph), loop-specific information
(e.g., loop iteration count, average loop iteration count per invocation),
and function-specific information (e.g., the average number of times that an invocation of a function invokes another).

\paragraph{Scheduler (SCD).}
NOELLE provides the \emph{scheduler} abstraction that offers the capability of moving instructions within and among basic blocks while preserving the original code semantics.
The scheduler relies on the PDG abstraction to guarantee semantic preservation.
The abstraction provides a hierarchy of schedulers starting from a generic one and including loop-specific and within-basic-block schedulers.
Each scheduler augments the generic capabilities with specialized capabilities (e.g., reducing the header size of a loop).

\paragraph{Loop Builder (LB).}
NOELLE offers the \emph{loop builder} abstraction that enables passes to modify/create/delete loops.
LB is similar to the IRBuilder abstraction offered by LLVM, but instead
of targeting instructions, LB targets loops.

\paragraph{Induction variables (IV).}
NOELLE provides the \emph{induction variable} abstraction.
Because LLVM's IR is in SSA form, the concept of the loop's induction variable is embodied by an SCC of the aSCCDAG of that loop.
NOELLE's abstraction exposes such SCC, the starting and ending value of an induction variable, the step amount per loop iteration, and whether an induction variable controls the number of loop iterations that will be executed.
We call governing induction variables those that control the number of loop iterations.
Finally, IV exposes the potential relationship with other induction variables for those that are derived.

The main difference between LLVM's induction variable and NOELLE's version is that the former only provides the PHI instruction that composes the SCC of an induction variable that belongs to the loop header.
Another difference is that NOELLE's version implements a more robust
algorithm to detect governing induction variables based on the aSCCDAG abstraction.
LLVM's implementation, instead, relies on the low-level def-use chains of the IR because of the lack of the SCCDAG abstraction within LLVM.
NOELLE's IV, therefore, detects more governing induction variables.

\paragraph{Induction Variable Stepper (IVS).}
A common operation for modern and emerging code transformations is to modify the step of induction variables.
For example, loop rotation needs to revert the step value of induction variables.
Another example is an advanced DOALL parallelization, which needs to
perform chunking between iterations to increase spatial locality.
The NOELLE's abstraction \emph{induction variable stepper} offers the
capability to modify any step value of induction variables of a loop;
users only need to specify the new step values, and the abstraction modifies the loop accordingly.

\paragraph{Loop (L).}
This abstraction includes a representation of the loop structure (called LS).
The latter is equivalent to the loop abstraction of LLVM.
The abstraction L, instead, adds to LS the loop dependence graph (computed from the PDG) and the loop-specific instances of the abstractions IV and INV.

\paragraph{Other abstractions.}
Above, we have described the most important abstractions NOELLE provides.
However, NOELLE provides additional abstractions used for simple
compilation tasks such as \emph{control equivalence}, \emph{reduction
operations}, \emph{extendible metadata} attached to control structures like
loops, \emph{SCCDAG partitioner}, \emph{forests(FR), and graphs} designed to restore connections among remaining parts when a node is deleted, \emph{architecture} to describe how logical cores are mapped to physical cores and NUMA nodes, and deterministic \emph{IDs} for instructions, loops, functions, and basic blocks.

Furthermore, NOELLE offers a new implementation of the \emph{loop
structure (LS)}, \emph{dominator}, and \emph{scalar evolution} abstractions.
This is because the LLVM abstractions computed by Function passes free their memory when they are invoked to analyze a different function.
This generates subtle, but unfortunately common, bugs that affect module passes.
The bug is generated when a module pass caches the pointers of the
abstractions returned by a function pass applied to multiple functions. All previous pointers but the last one are invalid.
This problem can become even more subtle because a function pass's
invocation can invalidate the abstraction returned by another function pass.
To avoid this common bug, NOELLE offers implementations of these LLVM
abstractions with the property that only their users can free these memory objects.

\subsection{NOELLE's Tools}
\label{sec:NOELLE_tools}
NOELLE includes tools (Table~\ref{tab:NOELLE_tools}) to help users deploy their compilation tool-chain
Next are the most important ones.

\textbf{\texttt{noelle-whole-IR}} generates a single IR file.
Merging all bitcode into a single bitcode file is important for the analyses and transformations that span a wide code region (e.g., the whole program).
Such an example is the alias analyses used by NOELLE to compute the PDG.
This tool is based on \texttt{gllvm}.

\textbf{\texttt{noelle-rm-lc-dependences}} modifies an IR program to remove
or reduce the impact of loop-carried data dependences (e.g., using Loop Builder to split a loop).

\textbf{\texttt{noelle-prof-coverage}} profiles IR code using representative program's inputs.
At the moment, NOELLE includes an instruction profiler, a branch profiler, and a loop profiler.

\textbf{\texttt{noelle-meta-pdg-embed}} computes the PDG of an IR file.
This tool computes the PDG by invoking many time-consuming and accurate alias analyses that power NOELLE.
Then, this tool embeds the computed PDG as metadata into the IR file so that NOELLE can re-construct the requested abstractions without requiring memory analyses.
This tool relies on NOELLE's PDG and IDs abstractions.

\textbf{\texttt{noelle-load}} loads the NOELLE's layer in memory.
Custom tools invoke NOELLE's empowered LLVM pass by using \texttt{noelle-load} rather than the LLVM tool \texttt{opt}.

\textbf{\texttt{noelle-arch}} measures architecture-specific characteristics.
At the moment, this tool measures the core-to-core latency and bandwidth.
This tool also interacts with the tool \texttt{hwloc}~\cite{hwloc} to find the number of physical and logical cores of the underlying platform, their mapping, and NUMA nodes.

\subsection{NOELLE's Testing}
\label{sec:NOELLE_testing}

NOELLE provides a testing infrastructure composed of hundreds of regression tests, unit tests, and performance tests.
These tests are micro C/C++ programs to illustrate corner cases or common code patterns found in popular benchmark suites such as SPEC CPU2017 and PARSEC 3.0.
This testing infrastructure allows NOELLE's users to quickly test their work with representative code patterns without paying the high compilation and profiling costs of the original codebase of the mentioned benchmark suites.
Finally, this infrastructure is integrated with distributed systems, such as HTCondor and Slurm, to run tests in parallel across multiple machines.
Optionally, NOELLE generates a bash file that executes all tests sequentially.

Tests are enabled by exposing NOELLE options and can be extended.
This allows NOELLE's users to surgically generate tests that stress a specific aspect of a specific code transformation.
For example, a user can force a parallelizing custom tool to parallelize only a given loop.

\noindent
\begin{table}[t!]
    \begin{minipage}{\columnwidth}
\input{Figs/alg_inv_llvm}
    \end{minipage} \\
    \begin{minipage}{\columnwidth}
\input{Figs/alg_inv_noelle}
    \end{minipage}
    \vspace{-1.5em}
\end{table}

\subsection{Impact of NOELLE's Abstractions}

NOELLE's abstractions may depend on each other to simplify design while
keeping high precision.
For example, the invariant abstraction (INV) uses the PDG
abstraction to identify loop invariants.
Next, we compare this NOELLE's implementation with the LLVM's to
highlight the impact of building upon higher-level abstractions
rather than lower-level ones.

Algorithm~\ref{alg:isInvariant-llvm} shows the simplified logic of LLVM's
implementation that relies on low-level abstractions to decide whether a given instruction is a loop invariant.
First, the algorithm checks if any operand of \texttt{I} is defined within loop \texttt{L}.
If no operands are defined within \texttt{L}, it checks the type of the instruction \texttt{I}.
If \texttt{I} is a load instruction, it checks if
any other instruction of \texttt{L} can modify the same memory location
accessed by \texttt{I}.
If \texttt{I} is a store instruction, it checks if any memory use precedes \texttt{I} in \texttt{L}.
If not, it checks no memory invalidation happens if \texttt{I} would be hoisted outside the loop.
Finally, if \texttt{I} is a call instruction, it checks (i) if \texttt{I} can modify any memory location, (ii) if the only memory accessed are via arguments to the call, (iii) and if any sub-loop can modify the same memory accessed via arguments by the call \texttt{I}.

Algorithm~\ref{alg:isInvariant-noelle} shows the NOELLE's implementation
that relies on the high-level PDG abstraction.
It checks if \texttt{I} is currently under analysis (i.e., in the stack \texttt{s}).
If not, it checks instruction that \texttt{I}
depends on whether it is outside the loop or a loop invariant.
Notice that this algorithm is smaller, simpler, and more precise than Algorithm~\ref{alg:isInvariant-llvm} (Figure~\ref{fig:All_invs}).

%% file: Tables/NOELLE_abstractions.tex
\begin{table*}[t]
\centering
\caption{
Abstractions provided by NOELLE
}
\vspace{-1em}
\label{tab:NOELLE_abs}
\resizebox{\linewidth}{!}{
\begin{tabular}{|l|l|r|l|}
\hline
\textit{Abstraction}    & \textit{Description}                                                                         & \textit{LoC}  & \textit{Depends on} \\

\hline
\hline
PDG                     & All dependences between instructions of a program                                                & 6775      &    \\
\hline                                                                                                                                  
aSCCDAG                 & SCCDAG of a loop with attributes on each SCC                                                     &           & PDG   \\
                        & (e.g., an SCC has loop-carried data dependence, it is reducible)                                 & 4517      &    \\
\hline                                                                                                                                  
Call graph (CG)         & Complete call graph of a program including indirect calls and their possible callees             & 620       & PDG   \\
\hline                                                                                                                                  
Environment (ENV)       & Variables needed by a task to execute (live-ins and live-outs)                                   & 991       & PDG   \\
\hline                                                                                                                                  
Task (T)                & Code region (and its inputs and outputs) executed by a thread                                    & 297       & ENV   \\
\hline                                                                                                                                  
Data-flow engine (DFE)  & Optimized engine to quickly evaluate data flow equations provided as inputs                      & 332       &  \\
\hline                      
Loop structure (LS)     & Describe the structure of a loop, its exits, latches, header, pre-header, basic blocks.          & 301       & \\
\hline
Profiler (PRO)          & Set of profilers at the IR level                                                                 & 1625      & LS  \\
\hline                                                                                                                                  
Scheduler (SCD)         & Mechanisms to change the schedule of instructions wthin and between basic blocks                 & 1523      & PDG, LS, DFE  \\
\hline                                                                                                                                  
Invariant (INV)         & Instructions, values, or memory locations that are loop invariants for a given loop              &  137      & PDG, LS       \\
\hline                                                                                                                                  
Induction variable      & Induction variables of a loop including the identification                                       &           & LS, INV     \\
(IV)                    & of the governing one (if it exists)                                                              & 352       & aSCCDAG     \\
\hline                                                                                                                                  
Induction variable      & Modifies the code of a loop to implement a change                                                &           & LS, INV, IV \\
stepper (IVS)           & in step value of its induction variables                                                         & 425       &            \\
\hline                                                                                                                                  
Reduction (RD)          & Identification and capability of reducing variables of a loop                                    & 868       & aSCCDAG, INV, IV    \\
\hline                                                                                                                                  
Loop (L)                & Canonical loop with its dependence graph, its SCCDAG, its invariants,                            &           & LS, PDG, IV,       \\
                        & its induction variables, and its exits                                                           & 1508      & INV, aSCCDAG, RD    \\
\hline                                                                                                                                  
Forest (FR)             & Forest of trees with the capability to adjust when a node is deleted to keep the connections     &           & L, CG           \\
                        & between the parent and the children of the deleted node                                          & 202       &                 \\
\hline                                                                                                                                  
Loop builder (LB)       & Set of loop transformations that modify a loop                                                   &           & FR, L, DFE,  \\
                        & (e.g., split a loop, translate do-while loops to while form and vice versa)                      & 4535      & IV, IVS, INV \\
\hline                                                                                                                                  
Islands (ISL)           & Capability to identify the disconnected sub-graphs of a graph (e.g., call graph, PDG)            &   56      & PDG, CG         \\
\hline                                                                                                                                  
Architecture (AR)       & Description of the underlying architecture in terms of logical/phisical cores, NUMA nodes.       &           &      \\
                        & It also provides the measured latencies and bandwidths between pairs of cores                    & 381       &      \\
\hline                                                                                                                                  
Others                  &                                                                                                  & 691       &    \\
\hline 
\hline 

                        & \textbf{LoC of NOELLE's abstractions } & \textbf{26142} & \\
\hline

\end{tabular}
}
\end{table*}

%% file: Tables/NOELLE_tools.tex
\begin{table*}[t]
\centering
\caption{
NOELLE's tools
}
\vspace{-1em}
\label{tab:NOELLE_tools}
\resizebox{\linewidth}{!}{
\begin{tabular}{|l|l|r|l|}
\hline
\textit{Tool}                                    & \textit{Description}                                                                                                         & \textit{LoC}  & \textit{Depends on}  \\

\hline
\hline
\texttt{noelle-whole-IR}            & Generate a single IR file from C/C++ source files embedding the compilation & 1522 &  \\
                                    & options as metadata inside the generated IR file                            &      &  \\
\hline
\texttt{noelle-rm-lc-dependences}   & Transform loops to remove as many loop-carried data dependences as possible & 912 & aSCCDAG, CG, \\
                                    &                                                                             &     & L, PRO, FR, LB \\
\hline                             
\texttt{noelle-prof-coverage}       & Inject code into the IR file given as input to profile IR instructions      & 1761 & PRO, FR, CG   \\
\hline         
\texttt{noelle-meta-prof-embed}     & Embed profiles into the IR file given as input                              & 152  & PRO, FR, CG \\
\hline
\texttt{noelle-meta-pdg-embed}      & Compute and embed the PDG into the IR file given as input                   & 451  & PDG \\
\hline
\texttt{noelle-load}                & Load the NOELLE abstractions into memory without computing them             & 12   & \\
\hline
\texttt{noelle-arch}                & Generate a file that describes the underlying architecture and its profiles (e.g., core-to-core latencies) & 259 & AR   \\
\hline                             
\texttt{noelle-linker}              & Links IR files together while preservering the semantic of metadata generated by NOELLE's tools & 59  & \\
\hline
\texttt{noelle-bin}                 & Generate a standalone binary from an IR file using the compilation options specified & 15 &  \\
                                    & as metadata inside the IR file given as input &  &  \\
\hline
\hline
                                    & \textbf{LoC of NOELLE's tools} & \textbf{5143} & \\
\hline
\end{tabular}
}
\end{table*}

%% file: Figs/alg_inv_llvm.tex
\setlength{\textfloatsep}{1em}
\begingroup
\removelatexerror
\begin{algorithm}[H]
    \newcommand*{\Let}[2]{ #1 $\gets$ #2}



    \scriptsize
    \KwResult{Return true if instruction I is an invariant in loop L}
    \tcc{Simplified logic of LLVM implementation} 
    \For{operand \KwSty{in} I.\FuncSty{getOperands}()}{
        \lIf{operand is defined in L}{\Return{False}}
    }

    \If{\FuncSty{isa<LoadInst>}(I)}{
        \For{Instruction J \KwSty{in} L}{
            \lIf{\FuncSty{getModRef}(\ArgSty{J}, \ArgSty{I}) != NoMod}{\Return{False}}
        }
    }

    \If{\FuncSty{isa<StoreInst>}(I)}{
        \For{memory use MU \KwSty{in} L}{
            \tcp{Conservatively ensures no memory\\use precedes this store}
            \lIf{\KwSty{not} DT.\FuncSty{dominates}(\ArgSty{I}, \ArgSty{MU})}{\Return{False}}
        }
        \tcp{Ensures no memory def/use would be\\invalidated by hoisting the store}
        \Let{M}{AA.\FuncSty{getNearestDominatingMemoryAccess}(\ArgSty{I})}\;
        \lIf{M is in L}{\Return{False}}

    }

    \If{\Let{call}{\FuncSty{dyn\_cast<CallInst>}(\ArgSty{I})}}{
        \lIf{AA.\FuncSty{getModRefBehavior}(\ArgSty{call}) != NoMod}{\Return{False}}
        \Let{S}{AA.\FuncSty{onlyMemoryAccessesAreArguments}(\ArgSty{call})}\;
        \lIf{\KwSty{not} S}{\Return{False}}
        \For{Argument A \KwSty{of} call}{
            \For{sL \KwSty{in} L->\FuncSty{getSubLoops}()}{
                \For{sI \KwSty{in} sL}{
                    \lIf{AA.\FuncSty{getModRef}(\ArgSty{A}, \ArgSty{sI}) != NoMod}{\Return{False}}
                }
            }
        }
    }

    \Return{True}\;
    \caption{\label{alg:isInvariant-llvm}\textit{isInvariant\_llvm(Instruction
    I, Loop L, \rightline{Dominator DT, AliasAnalysis AA)}}}
\end{algorithm}
\endgroup

%% file: Figs/alg_inv_noelle.tex
\setlength{\textfloatsep}{1em}
\begingroup
\removelatexerror
\begin{algorithm}[H]
\newcommand*{\Let}[2]{ #1 $\gets$ #2}
    \scriptsize
    \KwResult{Return true if instruction I is an invariant in loop L}
    \tcc{Implementation using high level abstraction PDG instead of low level abstractions alias analysis and dominators}
    \lIf{I \KwSty{in} s}{\Return{False}}
    s.\FuncSty{push}(\ArgSty{I})\;
    \For{PDG dependence J to I}{
        \If{J is in L}{
            \Let{\ArgSty{inv}}{\FuncSty{isInvariant\_noelle}(\ArgSty{J}, \ArgSty{L}, \ArgSty{dg}, \ArgSty{s})}\;
            \lIf{\KwSty{not} inv}{\Return{False}}
        }
    }
    s.\FuncSty{pop}()\;
    \Return{True};

    \caption{\label{alg:isInvariant-noelle}\textit{isInvariant\_noelle(Instruction I, Loop L, \rightline{PDG dg, Stack s})}}
\end{algorithm}
\endgroup

%% file: Sections/clients.tex
This section describes the code transformations built upon NOELLE.
Table~\ref{tab:NOELLE_clients} summarizes them and their LoC.

Each transformation relies on several of NOELLE's abstractions.
Table~\ref{tab:client_abstractions} shows the abstractions used by them. It is
important to notice that every abstraction is used by more than one custom tool
suggesting their wide applicability.

\paragraph{HELIX} parallelizes a loop by distributing its iterations between
cores~\cite{Campanoni:2012:HAP:2259016.2259028,Murphy:2016:PIT:2892208.2892214,6241522}. Each iteration is sliced into
several sequential and parallel segments. Different instances of the same
static sequential segment run sequentially between the cores while everything
else can overlap.

HELIX uses PRO, FR, and L of NOELLE to identify the most profitable loops to
parallelize. HELIX uses the PDG and ENV to identify and organize the live-in
and live-out of each chosen loop. LB and T abstractions are then used to
generate the parallel version of a loop.

HELIX uses aSCCDAG, INV, IV, and the RD abstractions to identify the SCCs that
need to be executed sequentially. HELIX uses DFE to translate SCCs into
sequential segments. SCD is then used to reduce the size of each sequential
segment as well as to schedule them within the body of each parallelized loop.
Moreover, HELIX uses IVS to perform chucking of loop iterations. Finally, HELIX
uses AR to implement helper thread
optimization~\cite{Campanoni:2012:HAP:2259016.2259028}.

\paragraph{DSWP} parallelizes a loop by distributing its SCCs between
cores~\cite{1540952}. Instances of a given SCC are executed by the same core to
create a unidirectional communication between cores. DSWP uses NOELLE's
abstractions, similarly to how HELIX does while leveraging DSWP-specific
knowledge to select the loops to parallelize and to parallelize them.

\paragraph{CARAT} is co-designed with the underlying operating system to
replace virtual memory. This compiler injects code to guard IR memory
instructions that cannot be proved at compile time to be
valid~\cite{10.1145/3385412.3385987}.

CARAT relies on the PDG, the aSCCDAG, and INV to identify the memory
instructions that need guarding. Then, it uses DFE and PRO to avoid redundant
guards of the same memory location. CARAT also uses L, LB, and IV to merge
guards. Finally, SCD is used to place the guards in the code.

\paragraph{Compiler-Based Timing} is co-designed with the underlying operating
system to inject calls to OS routines~\cite{COOS} into a program. This compiler
uses DFE and PRO to implement its specialized data flow analyses. It also uses
L, FR, and LB to handle potentially-infinite loops. Finally, it uses CG to
improve the accuracy of its time analyses.

\input{Tables/NOELLE_clients}

\paragraph{PRVJeeves} selects the pseudo-random value generators (PRVG) for a
randomized program (e.g., Monte Carlo
simulations)~\cite{10.1145/3368826.3377906}. To do so, it uses the PDG, CG, and
DFE to identify the allocations and uses of the PRVGs. Then, PRVJeeves uses PRO
to prune the design space (e.g., PRVGs not used frequently are left
unmodified). Moreover, it uses L, LB, INV, and IV to identify the uses of a
vector of PRVGs. Finally, PRVJeeves uses SCD to place the uses of a PRVG in the
code.

\paragraph{DOALL} parallelizes a loop that has no loop-carried data dependences
by distributing its iterations among cores~\cite{hurson1997parallelization}.
DOALL's implementation uses NOELLE's abstractions similarly to the other
parallelizing compilers (DSWP and HELIX), the difference being the loop
selection process and parts of the parallelized code generation. Yet, the loop
selection process and parts of the parallelized-code generation are naturally
different from the other parallelization techniques.

\paragraph{Loop Invariant Code Motion} hoists loop invariants outside their
loop. It uses FR to hoist loop invariants from innermost loops to outermost
ones. Then, it uses INV to identify instructions that could be hoisted.
Finally, it uses LB to perform the hoist transformation.

\paragraph{Time-Squeezer} generates code optimized for timing speculative
micro-architectures~\cite{DBLP:conf/isca/FanCJ19,
Fan:2018:CIC:3195970.3196013}. To this end, the compiler needs to decide when
to swap the compare operands (and modify its uses), how to change the schedule
of instructions, and where to inject instructions that modify the clock period
of the underlying architecture. This custom tool uses DFE, L, and FR to decide
where to inject clock-changing instructions. It then uses SCD to optimize the
instruction sequence of a code region that uses the same clock period
previously chosen. Finally, it uses ISL and PDG to analyze the compare
instructions and their dependences.

%% file: Tables/NOELLE_clients.tex
\begin{table*}[t]
\centering
\caption{
Custom tools built upon NOELLE
}
\vspace{-1em}
\label{tab:NOELLE_clients}
\resizebox{\linewidth}{!}{
\begin{tabular}{|l|l|r|r|r|}
\hline
                                  &                                                                                                  &                   & \textit{LLVM +}   & \textit{Percent}   \\
\textit{Custom tool}              & \textit{Description}                                                                             & \textit{LLVM}     & \textit{NOELLE}   & \textit{reduction} \\

\hline
\hline
Time Squeezer (TIME)              & Compiler to optimize compare instructions for timing speculative architectures                   &   510         &            92     &  82.0\%  \\
\hline                                                                                                                                                                    
Compiler-based timing (COOS)      & Compiler to inject calls to Operating System routines to replace hardware interrupts             &  1641         &           495     &  69.8\%  \\
\hline                                                                                                                                                                    
Loop Invariant Code Motion (LICM) & Hoist loop invariants outside their loop                                                         &  2317         &           170     &  92.7\%  \\
\hline                                                                                                                                                                    
DOALL                             & Parallelizing compiler that applies the DOALL code parallelization technique                     &  5512         &           321     &  94.2\%  \\
\hline                                                                                                                                                                    
Dead Function Elimination (DEAD)  & Reduce the number of functions without increasing the total number of IR instructions            &  7512         &            61     &  99.2\%  \\
\hline                                                                                                                                                                    
DSWP                              & Parallelizing compiler that applies the DSWP code parallelization technique                      &  8525         &           775     &  90.9\%  \\
\hline                                                                                                                                                                    
HELIX                             & Parallelizing compiler that applies the HELIX code parallelization technique                     & 15453         &           958     &  93.8\%  \\
\hline                                                                                                                                                                    
PRVJeeves (PRVJ)                  & Compiler to select the Pseudo Random Value Generators for the program given as input             & 17863         &           456     &  97.4\%  \\
\hline                                                                                                                                                                    
CARAT                             & Inject memory guards to potentially incorrect memory instructions                                & 21899         &           595     &  97.3\%  \\
\hline                                                                                                                                                                    
Perspective (PERS)                & Parallelizing compiler that minimizes speculation and privatization costs                        &  33998        &         22706     &  33.2\%  \\
\hline
\end{tabular}
}
\end{table*}

%% file: Sections/evaluation.tex
This section presents evaluation results for NOELLE and the custom tools built
upon NOELLE. Before presenting the results, we first describe our evaluation
platform and our evaluation methodology. Our results show that each NOELLE's
abstraction can be used by several and significantly different custom tools.
Results suggest that NOELLE's implementation of a few abstractions that exist
in LLVM is more precise than their LLVM counterparts. Finally, results suggest
that we can build a custom tool in a few lines of code that is powerful enough
to improve the performance or reduce the binary size compared to the mainline,
wildly adopted compilers like \texttt{clang}.

\input{Tables/clients_abstractions}

\subsection{Experimental Setup}
We have evaluated NOELLE and ten custom tools on the platform described next and by following the measurement methodology described here.

\label{sec:evaluation-abstractions}
\begin{figure*}[t]
  \centering
  \includegraphics[width=\textwidth]{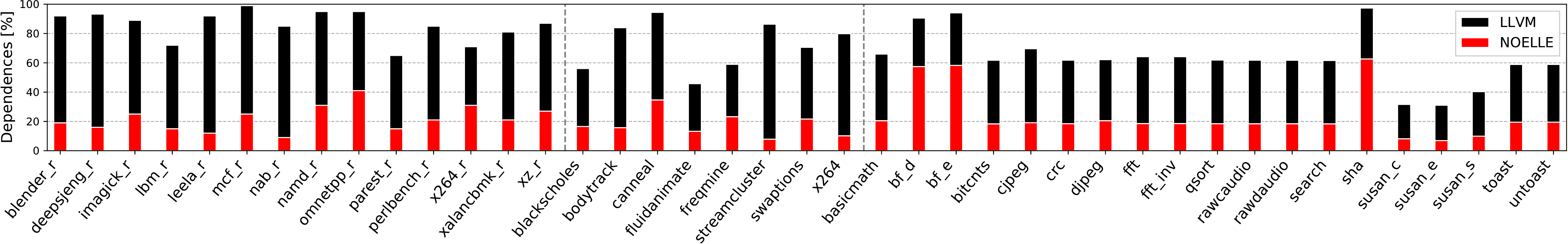}
  \vspace*{-5mm}
  \caption{
  \label{fig:All_deps}
  \small{
While LLVM is capable of proving the non-existence of most dependences,
NOELLE disproves more by relying on state-of-the-art alias analysis
techniques (SCAF~\cite{apostolakis:20:pldi})
\vspace{-0.3cm}
  }
}
\end{figure*}

\textbf{Platform. } Our evaluation was done on a Dell PowerEdge R730 server
with one Intel Xeon E5-2695 v3 Haswell processor running at 2.3GHz. The
processor has 12 cores with 2-way hyperthreading, 35MB of last-level cache, and
has a peak power consumption of 120W. The cores are supported by 256GB of main
memory in 16 dual rank RDIMMs at 2133MHz. Turbo Boost was disabled, and no CPU
frequency governors were used (i.e., all cores ran at a maximum frequency). The
OS used is Red Hat Enterprise Linux Server 8 on kernel 4.18. NOELLE was built
on top of LLVM 9~\cite{lattner2004llvm}.

\textbf{Statistics and convergence.} Each data point we show in our evaluation
is an average of multiple runs. We ran the relevant configurations as many
times as necessary to achieve a tight confidence interval (95\% of the measurements are within 5\% of the mean).
 \subsection{Building Upon NOELLE Reduces Source Code}
\label{sec:evaluation-reduces} NOELLE simplifies the implementation of code
analyses and transformations. Table~\ref{tab:NOELLE_clients} compares the
implementations of 10 transformations when built upon NOELLE and when
implemented only using LLVM abstractions. The reduction in LoC is significant,
reducing the maintainability cost of these custom tools.

NOELLE abstractions are general enough to be useful by many and highly
heterogeneous custom tools. Table~\ref{tab:client_abstractions} shows that each
abstraction is used by several custom tools. For example, the loop builder (LB)
is used by eight custom tools out of 10. Moreover, it is important to notice
the heterogeneity of these custom tools that use (for example) LB:
parallelizing transformations, loop invariant code motion (LICM), compare
instruction optimization and code generation for timing speculative
micro-architecture (TIME), memory guard injector and optimization (CARAT), PRVG
selector (PRVJ), and scheduler of OS routines within applications (COOS).

\subsection{NOELLE Abstractions}

\begin{figure*}[t]
  \centering
  \includegraphics[width=\textwidth]{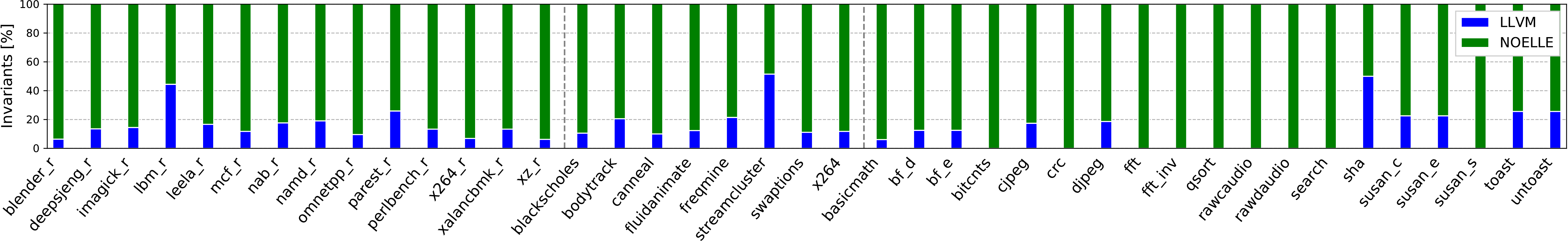}
  \vspace*{-5mm}
  \caption{
  \label{fig:All_invs}
  \small{
      NOELLE detects significantly more invariants than LLVM even if the former
      relies on a simpler and shorter algorithm powered by higher-level
      abstraction (Algorhtm~\ref{alg:isInvariant-noelle}) compared to LLVM
      (Algorithm~\ref{alg:isInvariant-llvm}).
  }
}
\end{figure*}

Next, we compare the subset of NOELLE's abstractions that are also available in
LLVM. These abstractions are loop invariants, loop induction variables, and
dependences.

Figure~\ref{fig:All_deps} shows that NOELLE's implementation of dependences
within the PDG abstraction is more accurate than LLVM's abstraction. LLVM is
capable of proving a significant fraction of potential memory dependences
non-existing. NOELLE further improves these results dramatically by leveraging
state-of-the-art alias
analyses~\cite{apostolakis:20:pldi,Johnson:2017:CDA:3049832.3049849,sui2016svf}.

Figure~\ref{fig:All_invs} compares the number of loop invariants identified by
both LLVM and NOELLE. NOELLE identifies significantly more loop invariants than
LLVM because the invariant abstraction of NOELLE is built using the PDG
abstraction. This makes the invariant detection algorithm within NOELLE
(Algorhtm~\ref{alg:isInvariant-noelle}) smaller, more elegant, and more
powerful compared to the LLVM one (Algorithm~\ref{alg:isInvariant-llvm}).

Finally, we computed the number of loop induction variables that govern a loop
using both LLVM and NOELLE. We did so for the three benchmark suites for a
total of 41 benchmarks. LLVM identifies only a few loop induction variables (11
total) among all loops for the 41 benchmarks. The reason is that LLVM's
induction variable analysis expects the input IR to have loops in the do-while
shape. However, most loops in the 41 benchmarks have a while shape, and
changing them into a do-while shape would reduce the applicability of all the
implemented parallelization techniques. Instead, NOELLE identifies many loop
induction variables (385 total) independently of the shape of the loop being
analyzed.

\subsection{Parallelizing Transformations Upon NOELLE}

\begin{figure*}[htp]
  \centering
  \includegraphics[width=\textwidth]{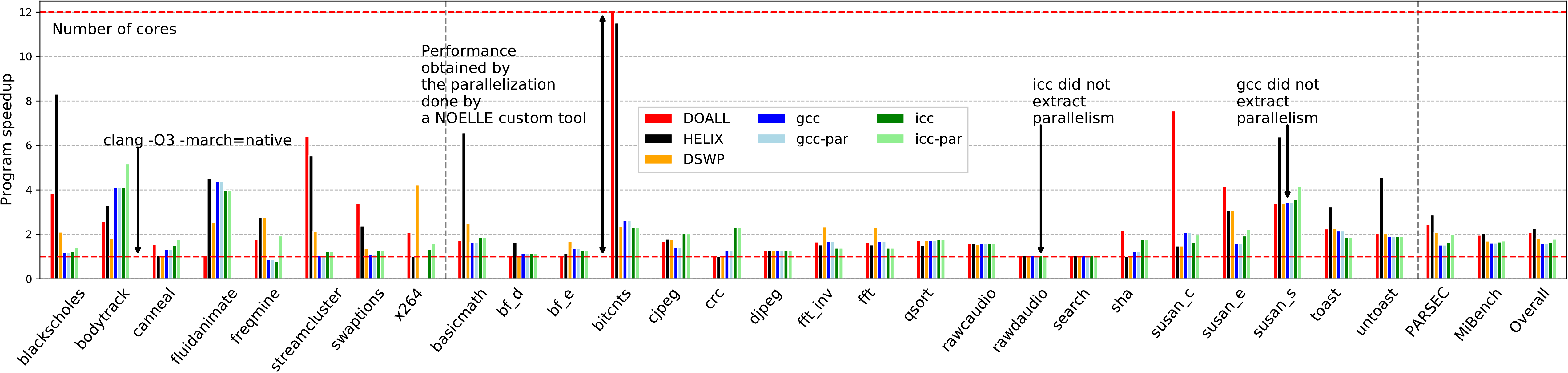}
  \vspace*{-5mm}
  \caption{
  \label{fig:All_speedup}
  \small{
      Both \texttt{gcc} and \texttt{icc} did not obtain additional performance
      benefits from their parallelization techniques.  Instead, NOELLE-based
      parallelizing tools generate additional benefits compared to their
      baseline, \texttt{clang}.
\vspace{-0.3cm}
  }
}
\end{figure*}

Next, we describe the parallelizing code transformations built upon NOELLE
(HELIX, DSWP, DOALL) that do not rely on speculative techniques. This allows us
to compare few-hundred lines of code implementations built upon NOELLE with the
parallelizing transformations implemented by \texttt{icc} (Intel) and
\texttt{gcc} (GNU) compilers.

Figure~\ref{fig:All_speedup} shows the speedups we obtained in PARSEC and
MiBench benchmark suites. The few missing benchmarks have failed to compile
with the unmodified \texttt{clang} compiler, and therefore we cannot use them
to test NOELLE-based tools. Figure~\ref{fig:All_speedup} shows that the
NOELLE-based small custom tools already extract more parallelism compared to
what \texttt{gcc} and \texttt{icc} extract. Furthermore, we analyzed the few
benchmarks that NOELLE-based parallelizing tools could not extract significant
performance benefits (e.g., \texttt{crc}). We found this is due to the lack of
support for memory object cloning. This is arguably an abstraction that should
exist in the parallelization techniques rather than within NOELLE as the latter
is not specialized for parallelization purposes.

We also run these five parallelizing tools on 14 SPEC CPU2017 benchmarks (the
only missing benchmark is \texttt{gcc}, which did not compile with
\texttt{clang}). Speedups were obtained only by NOELLE-based parallelizing
tools and are within 1\% and 5\% for these 14 benchmarks demonstrating the
robustness of NOELLE abstractions.  Speculative techniques are likely to be
required to unlock further speedups on these benchmarks. We argue that
speculative techniques should be implemented outside NOELLE as they are
specific to the parallelization goal.

Finally, we have ported a state-of-the-art parallelizing compiler
(Perspective~\cite{apostolakis:20:asplos}) together with the authors. We
modified the original codebase to use the PDG and the aSCCDAG abstractions.
This new version has preserved the performance shown in the authors' original
paper.

\subsection{Reducing Binary Size with NOELLE}

Binary size is an important optimization goal for both embedded systems and
servers~\cite{10.1145/3307650.3322234}. The compiler \texttt{clang} offers an
optimization level specialized for this goal (\texttt{-Oz}).
DeadFunctionElimination further reduces the binary size by 6.3\% on average
among the 41 benchmarks considered.

%% file: Tables/clients_abstractions.tex
\begin{table}
\centering
\caption{
Each NOELLE's abstraction is used by several custom tools.
}
\vspace{-1em}
\label{tab:client_abstractions}
\resizebox{1\linewidth}{!}{
\begin{tabular}{|l|c|c|c|c|c|c|c|c|c|c|c|c|c|c|c|c|c|c|}
\hline
 \begin{tabular}[c]{@{}l@{}}\textbf{Custom}\\ \textbf{tool}\end{tabular}& \multicolumn{18}{c|}{\textbf{NOELLE's abstractions used}}                                                                                                                            \\ \cline{2-19}
  & \rotatebox{90}{{\Small PDG}}    & \rotatebox{90}{{\Small aSCCDAG }} & \rotatebox{90}{{\Small CG}}     & \rotatebox{90}{{\Small ENV}}         & \rotatebox{90}{{\Small T}}      & \rotatebox{90}{{\Small DFE}}       & \rotatebox{90}{{\Small PRO}}      & \rotatebox{90}{{\Small SCD}}       & \rotatebox{90}{{\Small L}}      & \rotatebox{90}{{\Small LB}}      & \rotatebox{90}{{\Small IV}}        & \rotatebox{90}{{\Small IVS}}       & \rotatebox{90}{{\Small INV}}        & \rotatebox{90}{{\Small FR}}     & \rotatebox{90}{{\Small ISL}}     & \rotatebox{90}{{\Small RD}}        & \rotatebox{90}{{\Small AR}}  & \rotatebox{90}{{\Small LS}}          \\
\hline
\hline
{\Small HELIX}           & \cmark & \cmark  &        & \cmark      & \cmark & \cmark    & \cmark   & \cmark    & \cmark & \cmark  & \cmark    & \cmark    & \cmark     & \cmark &         & \cmark    & \cmark   & \cmark     \\
\hline
{\Small DSWP}            & \cmark & \cmark  &        & \cmark      & \cmark &           & \cmark   & \cmark    & \cmark & \cmark  & \cmark    & \cmark    & \cmark     & \cmark &         & \cmark    & \cmark    & \cmark      \\
\hline
{\Small CARAT}           & \cmark & \cmark  &        &             &        & \cmark    & \cmark   & \cmark    & \cmark & \cmark  & \cmark    &           & \cmark     &        &         &           &      & \cmark           \\
\hline
{\Small COOS}            &        &         & \cmark &             &        & \cmark    & \cmark   &           & \cmark & \cmark  &           &           &            & \cmark &         &           &       & \cmark          \\
\hline
{\Small PRVJ}            & \cmark &         & \cmark &             &        & \cmark    & \cmark   & \cmark    & \cmark & \cmark  & \cmark    &           & \cmark     &        &         &           &       & \cmark          \\
\hline
{\Small DOALL}           & \cmark & \cmark  &        & \cmark      & \cmark &           & \cmark   &           & \cmark & \cmark  & \cmark    & \cmark    & \cmark     & \cmark &         & \cmark    & \cmark   & \cmark       \\
\hline
{\Small LICM}            &        &         &        &             &        &           &          &           & \cmark & \cmark  &           &           & \cmark     & \cmark &         &           &     & \cmark           \\
\hline
{\Small TIME}            & \cmark &         &        &             &        & \cmark    &          & \cmark    & \cmark & \cmark  &           &           &            & \cmark & \cmark  &           &       & \cmark         \\
\hline
{\Small DEAD}            &        &         & \cmark &             &        &           &          &           &        &         &           &           &            &        & \cmark  &           &       &       \\
\hline
{\Small PERS}            & \cmark & \cmark  &        &             &        &           &          &           &        &         &           &           &            &        &         &           &        &      \\
\hline
\end{tabular}
}
\end{table}

%% file: Sections/related_work.tex







\paragraph{Providing High-level Abstractions}

Researchers have explored bringing high-level abstractions to compilers
in many different ways.
A few compilers that support automatic parallelization, including
Polaris~\cite{blume::PolarisNextGeneration}, a parallelizing compiler for
Fortran programs, Cetus~\cite{::CetusProject}, a C compiler focusing on
multicore, ROSE~\cite{::RoseCompilerProgram}, a compiler for building custom
compilation tools, operate on high-level abstractions and perform
source-to-source translation, and thus miss opportunities presented only in
low-level IRs including more fine-grained operations and more canonical code
patterns.

\begin{sloppypar}
Many domain-specific projects add new abstractions similar to NOELLE.
SeaHorn~\cite{gurfinkel:2015:SeaHornVerificationFramework} provides new
abstractions for developing new verification techniques.
 Polly~\cite{::PollyPolyhedralOptimizations, grosser2011polly},
PLUTO~\cite{bondhugula2007pluto}, HALIDE~\cite{::Halide,
ragan-kelley:2013:HalideLanguageCompiler}, Tiramisu~\cite{::TiramisuCompiler,
baghdadi:2019:TiramisuPolyhedralCompiler}, and
APOLLO~\cite{caamano:2017:APOLLOAutomaticSpeculative} provide abstractions to
suit polyhedral optimizations, which target loops characterized by regular
control and data flows. TensorFlow~\cite{::TensorFlow}, a widely used machine
learning framework, uses high-level graph representations that allow graph
optimizations more discoverable~\cite{::TensorFlowGraphOptimization}. These
projects focus on specific domains and their abstractions are not easily reusable for problems outside their domains.
\end{sloppypar}

Few domain-independent compilers combine low-level IR with high-level
abstractions like NOELLE. SUIF compiler~\cite{::SUIFCompilerSUIF} provides
low-level IR as well as higher-level constructs including loops, conditional
statements, and array accesses~\cite{wilson::SUIFCompilerSystem}. The IMPACT
compiler~\cite{chong::IMPACTArchitecturalFramework}, which provides
hierarchical IRs to enable optimizations at different levels. Unfortunately,
they are not maintained anymore.

The LLVM community also has a Loop Optimization Working
Group~\cite{llvm::LoopOptimizationWorking} that recently has started working
on a few abstractions included in NOELLE, such as dependence graph. We plan
to share NOELLE code with them. We also see value in maintaining NOELLE as a
separate project that focuses mainly on performance rather than making a
balance between performance, code size, and compilation time.


\paragraph{LLVM Projects}
As we have built NOELLE on top of LLVM, we want to know how NOELLE would do.
To do this, we have exhaustively reviewed all 544 papers published in
PLDI, CGO, and CC during the past five years (2016-2020). Out of these
papers, 87 papers explicitly mention that they are built on top of LLVM by
either implementing new passes, modifying the LLVM internals, or creating a
new front-end/back-end based on LLVM IR.
Out of these 87 papers,

\begin{itemize}[leftmargin=*,topsep=0cm]
\item 26 (29.9\%) use abstractions similar to the ones provided by NOELLE.
Thus, they can potentially be re-implemented on top of NOELLE with significantly
fewer lines of code and/or with better performance. We have implemented
CARAT~\cite{10.1145/3385412.3385987} and
PRVJeeves~\cite{10.1145/3368826.3377906} in NOELLE and presented the benefit
in \ref{sec:clients}. Other examples include Spinal
Node~\cite{kim:2019:SpinalCodeAutomatic}, which uses PDG as well as data flow
analysis; Valence~\cite{zhou:2019:ValenceVariableLength}, which relies on call
graph analysis;
Clairvoyance~\cite{tran:2017:ClairvoyanceLookaheadCompiletime}, which relies
on loop-carried dependence analysis.

\item 10 (11.5\%) provide new abstractions or implement analyses or
transformations that fulfill NOELLE abstractions. We have already integrated
SVF~\cite{sui:2016:SVFInterproceduralStatic} and
SCAF~\cite{apostolakis:20:pldi} within NOELLE.
We plan to evaluate other
examples~\cite{doerfert:2017:OptimisticLoopOptimization,
manilov:2018:GeneralizedProfileguidedIterator,
maalej:2017:PointerDisambiguationStrict,
phulia:2020:OOElalaOrderofevaluationBased} in the future.

\item 25 (28.7\%) are doing tasks orthogonal to NOELLE's abstractions.
Nevertheless, they do not conflict with NOELLE because both implementations
do not modify LLVM internals. Due to NOELLE's modular and demand-driven
design, future work can use NOELLE even if only a subset of abstractions are
of interest.

\item 26 (29.9\%) paper modify LLVM internals or use alternative
front-end/back-end. These projects need to be analyzed case by case for
the possibility of integration with NOELLE.

\end{itemize}

In conclusion, 41.4\% of the projects are highly likely to benefit
from or contribute to NOELLE's abstractions; 28.7\% have the potential for future
collaboration; 29.9\% need investigation before integration.

%% file: Sections/conclusion.tex
Code analyses and transformations need to go beyond peephole and ILP
optimizations for modern architectures. Their implementation requires
high-level abstractions that are currently lacking in LLVM. This paper
introduces NOELLE, an open-source compilation layer built upon LLVM that
provides the required abstractions. NOELLE has been tested with ten highly
diverse and complex tools that are built upon it. All these tools gain benefits
compared to unmodified LLVM while dramatically reducing their LoC.